\documentclass{article}

\usepackage{amssymb,amsfonts,amsmath,stmaryrd}
\usepackage{cite,enumerate,float,indentfirst}
\usepackage{color}
\usepackage{tikz}
\usetikzlibrary{arrows,snakes,backgrounds}

\def\be{\begin{eqnarray}}
\def\ee{\end{eqnarray}}
\def\nn{\nonumber}

\def\tr{{\rm tr}\,}
\def\Tr{{\rm Tr}\,}

\def\S{{\rm Schur}}
\def\J{{\rm Jack}}
\def\M{{\rm Mac}}

\definecolor{red}{rgb}{1,0,0}
\definecolor{orange}{rgb}{1,0.5,0}
\definecolor{violet}{rgb}{0.7,0,1}

\hoffset= - 1.0in         

\begin{document}

\begin{center}
\begin{small}
\hfill MIPT/TH-01/22\\
\hfill FIAN/TD-01/22\\
\hfill ITEP/TH-02/22\\
\hfill IITP/TH-01/22\\

\end{small}
\end{center}

\vspace{.5cm}

\begin{center}
\begin{Large}\fontfamily{cmss}
\fontsize{15pt}{27pt}
\selectfont
	\textbf{Superintegrability summary}
	\end{Large}
	
\bigskip \bigskip

\begin{large}
A. Mironov$^{a,b,c,}$\footnote{mironov@lpi.ru; mironov@itep.ru},
A. Morozov$^{d,b,c,}$\footnote{morozov@itep.ru},
\end{large}

\bigskip

\begin{small}
$^a$ {\it Lebedev Physics Institute, Moscow 119991, Russia}\\
$^b$ {\it Institute for Information Transmission Problems, Moscow 127994, Russia}\\
$^c$ {\it ITEP, Moscow 117218, Russia}\\
$^d$ {\it MIPT, Dolgoprudny, 141701, Russia}
\end{small}
 \end{center}

\bigskip

\begin{abstract}
We enumerate generalizations of the superintegrability property
$<character>\ \sim {\rm character}$
and illuminate possible general structures behind them.
We collect variations  of original formulas
available up to date,
and emphasize the remaining difference between the cases of Hermitian and complex matrices,
bosonic and fermionic ones.
Especially important is that the story is in no way restricted
to Gaussian potentials.
\end{abstract}

\bigskip

\section{Introduction}

In \cite{MM}, we suggested to consider in detail
a remarkable property of the Gaussian matrix averages:
averages of Schur functions are again
expressed through the same Schur functions (see many examples in \cite{DiF}--\cite{MMMZh}, and also some preliminary results in \cite{Kaz}--\cite{MKR}).
The two basic examples were
\be
\Big<\S_R[ZZ^\dagger] \Big>:=\int \S_R[ZZ^\dagger] \,e^{-\tr ZZ^\dagger} d^2Z =
\frac{\S_R\{N_1\}\cdot \S_R\{N_2\}}{ \S_R\{\delta_{k,1}\}}
\label{siComp}
\ee
for the integral over complex $N_1\times N_2$ matrices $Z$ (rectangular complex model, RCM)
and
\be
\Big<\S_R[X] \Big>:=
\int \S_R[X] \,e^{-\frac{1}{2}\tr X^2} dX =\frac{\S_R\{N\}\cdot \S_R\{\delta_{k,2}\}}{ \S_R\{\delta_{k,1}\}}
\label{siHerm}
\ee
for the integral over Hermitian $N\times N$ matrices $X$ (Hermitian model, HM).
Here $\S_R[A]$ denotes the value of the Schur function $\S_R\{p_k\}$ as a graded polynomial of the power sums $p_k$
at the Miwa locus $p_k=\tr A^k$,
while $\S_R\{N\} = \S_R[I_N]$ corresponds  to $p_k=\tr I_N^k = N$
with a unit square matrix $I$ of the size $N$. {\bf Throughout this paper, all the integrals are normalized in such a way that
the average $<1>=1$.}
The quantity in the denominators ${ \S_R\{\delta_{k,1}\}}$ is often denoted $d_R$.
We denote all the averages by $\Big<\S_R \Big>$
to emphasize that the {\it meaning} of the procedure is always the same,
as is the statement, {\bf the average of a proper symmetric function is proportional to the same symmetric function at a peculiar locus},
but a concrete definition of the averaging procedure varies from case to case.

The Schur functions {\it per se} are explicitly defined
for any Young diagram $R = \{r_1\geq r_2 \geq\ldots \geq r_{l(R)}>0\}$
either by the determinant Jacobi-Trudy formula, or by the Frobenius formula \cite{Mac,Fulton},
but more important is conceptual definition as a character,
of a linear group,
and as a common set of eigenfunctions for all generalized cut-and-join operators $W_\Delta$ \cite{MMN},
\be
\hat W_\Delta \,\S_R = \phi_R(\Delta) \cdot \S_R
\ee
where $\phi_R(\Delta)$ is an adequate analytic continuation of a peculiarly normalized symmetric group
characters $\psi_R(\Delta)$ to arbitrary pairs of Young diagrams $R$ and $\Delta$,
perhaps, of distinct sizes.

In \cite{IMM}, we suggested to call the properties (\ref{siComp}) and (\ref{siHerm})
{\it superintegrability} because of the following analogy.
Partition functions of matrix models, i.e. averages
$\Big<\exp \sum_k p_k\tr X^k/k\Big> = \Big<\sum_R \S_R\{p_k\} \cdot \S_R[X]\Big>$,
are long known to be the $\tau$-functions of integrable hierarchies \cite{UFN3}.
However these are not generic $\tau$-functions, but satisfying a peculiar constraint:
{\it the string equation}, and, as a corollary, the entire set of Virasoro-like constraints.
Thus they are {\it more} than just integrable, they are explicitly comprehensible.
The situation looks like a far-going generalization of the well-know fact
that the motion in every potential $r^n$ is integrable, but the orbits are closed,
and the answers are expressible through elementary functions only for $n=2$ and $n=-1$
(harmonic oscillator and Coulomb potentials).
The fact that the orbits are closed can be explained as due to an additional conservation law,
it converts integrability to {\it super}integrability and makes the problem
exactly solvable.
We assume that  (\ref{siComp}) and (\ref{siHerm}) give us the same kind of signals,
this time about the existence of a peculiar basis, in which the averages can be
explicitly evaluated.

The purpose of this paper is to discuss generalizations of (\ref{siComp}) and (\ref{siHerm}),
which are already known in different directions to come a little closer to an understanding of this remarkable phenomenon.
Given the fact that the matrix models provide the best known elementary approximation
to generic string theory, one can hope that, in this way, we can learn something important
about strings, which could explain in which sense the string theory is distinguished
and comprehensible.

A simple picture illustrates what is currently known.
The original superintegrability relations (\ref{siComp}) and (\ref{siHerm})
can be lifted (refined) to (\ref{siCompAB1}) and (\ref{siHermA1}),
where the dependence on size of the matrix, $N=\tr I$ is promoted to that on an arbitrary matrix.
These refined formulas admit generalizations to tensor and fermion models,
but the latter case is different for the RCM and the HM, where the fermionic representation
is substituted by the Q-Schur functions. Note also that formulas for the tensor models are known so far only for extensions of the RCM, since they are much simpler in this case. One can also change the Vandermonde measure in the eigenvalue realizations and the Gaussian potential, but most results are currently known only in the non-refined case.

\be
\begin{array}{ccccccccc}
&&&&\text{fermionic} \ \ \ \ Q-\text{Schur} \\
&&&&\!\!\!(\ref{fermAB}) \ \ \ \ \ \ \ (\ref{QRA}) \\
\\
&&&&\uparrow \ \ \ \ \uparrow \\
\\
&&&&  (\ref{siCompAB1})+(\ref{siHermA1}) \\
{\rm tensor} \ (\ref{tensor}) &&\leftarrow && \uparrow \\
&&&& (\ref{siComp})+(\ref{siHerm}) \\
\\
&&&&  \ \ \ \downarrow &\!\!\! \searrow \\
\\
&&&& \text{deformed potential} && \!\!\!\text{deformed measure}\\
&&&& \swarrow \ \ \ \searrow   &&  \swarrow \ \ \ \   \searrow \\
&&&& \text{monomial} \ \ \ \ \ \text{logarithmic} && \ \ \ \ \ q,t \ \ \ \
 \text{trigonometric} \\
&&&& \!\!\!\!\! \text{sec.}\ref{monom} \ \ \ \ \ \ \text{sec.}\ref{log} &&
\!\!\!\! \!\!\!\!\!\!\!\! \!\!\!\! \text{sec.}\ref{qt} \ \ \ \ \phantom{AAA}
\\
\end{array}
\ee

\section{Refinement
\label{refinment}}

\subsection{The basic case}

A feature of (\ref{siComp}) and (\ref{siHerm}) that catches the eye is that, while the characters
at the l.h.s. depend on the non-trivial matrix, those at the r.h.s., on $N=\tr I$, the unit matrix rather than arbitrary.
It comes with no surprise that this restriction can be easily lifted.
The relevant generalizations are
\be
\int \S_R[ZZ^\dagger] \,e^{-\tr AZBZ^\dagger } d^2Z =
\frac{\S_R[A^{-1}]\cdot \S_R[B^{-1}]}{d_R}
\label{siCompAB1}
\ee
for the RCM  \cite{Mac,Ivan1,Ivan2,Orlov,NO,MMkon},
and
\be
\int \S_R[X] \,e^{-\frac{1}{2}\tr AXAX} dX =
\frac{\S_R[A^{-1}]\cdot \S_R\{\delta_{k,2}\}}{ \S_R\{\delta_{k,1}\}}
\label{siHermA1}
\ee
for the HM \cite{MMl}.
One can keep the Gaussian weights intact but change instead the arguments
of the Schur functions:
\be
\boxed{
\Big<\S_R[AZBZ^\dagger] \Big> =
\int \S_R[AZBZ^\dagger] \,e^{-\tr ZZ^\dagger } d^2Z =
\frac{\S_R[A ]\cdot \S_R[B ]}{ d_R}
}
\label{siCompAB2}
\ee
and
\be
\boxed{
\Big<\S_R[AX] \Big> =
\int \S_R[AX] \,e^{-\frac{1}{2}\tr X^2} dX =
\frac{  \S_R\{\delta_{k,2}\}\cdot \cdot \S_R[A ]}{ \S_R\{\delta_{k,1}\}}
}
\label{siHermA2}
\ee

\subsection{Tensor models}

One of the natural generalizations of the RCM is the theory of rank-$r$ tensors \cite{IMM,MMten},
which contains the complex $N_1\times\ldots\times N_r$ fields $Z_{a_1\ldots a_r}$
with the refined Gaussian action \cite{MMkon}
\be
S := \sum_{a_1,b_1=1}^{N_1} \ldots \sum_{a_r,b_r=1}^{N_r}
Z_{a_1\ldots a_r} \Big(Z^\dag\Big)^{b_1\ldots b_r} \prod_{i=1}^r \Big(A_{(i)}\Big)^{a_i}_{b_i}
\ee
The substitutes of the Schur functions (generalized tensor characters)
are $A$-independent combinations depending on $r$ different representations
of the same size $n=|R_1|=\ldots=|R_r|$,
\be
\chi_{R_1,\ldots,R_r}(Z, Z^\dag) :=
\frac{1}{n!}\sum_{\sigma_1,\ldots,\sigma_r \in S_n}
\psi_{R_1}(\sigma_1)\ldots \psi_{R_r}(\sigma_r)\cdot
{\cal O}_{\sigma_1,\ldots,\sigma_r}
\label{tensor}
\ee
where analogues of multi-trace operators
depend on $r$ permutations from the double coset $S_n\backslash S_n^{\otimes r}/S_n$:
\be
{\cal O}_{\sigma_1,\ldots,\sigma_r} = \sum_{\vec a^1=1}^{N_1}\ldots \sum_{\vec a^r=1}^{N_r}
\left(\prod_{p=1}^n Z_{a_p^1\ldots a_p^r}\Big(Z^\dag\Big)^{ a_{\sigma_1(p)}^1  \ldots a_{\sigma_{r(p)}}^r  }\right)
\ee
The Gaussian averages of these operators (which form a basis in the space of all operators with non-zero Gaussian averages) manifest the superintegrability property in the form
\be
\Big< \chi_{R_1,\ldots,R_r}\Big>
:=\int \chi_{R_1,\ldots,R_r}(Z,Z^\dag) e^Sd^2Z=
\ C_{R_1,\ldots,R_r} \cdot
{\S_{R_1}\left\{\Tr A_{(1)}^{-k}\right\}\over\S_{R_1}\{\delta_{k,1}\}}
\cdot \ldots\cdot {\S_{R_r}\left\{\Tr A_{(r)}^{-k}\right\}\over\S_{R_r}\{\delta_{k,1}\}}
\ee
where
\be
C_{R_1,\ldots,R_r}:=\sum_{\Delta\vdash n}{\prod_{i=1}^r \psi_{R_i}(\Delta)\over z_\Delta}
\ee
$\psi_R(\Delta$ is the character of the symmetric group $S_n$, and $z_\Delta$ is the standard symmetric factor of the Young diagram $\Delta$ (order of the automorphism) \cite{Fulton}.
In the case of $r=3$, $C_{R_1,R_2,R_3}$ are the Clebsch-Gordan coefficients of the three irreducible representations $R_1$, $R_2$, $R_3$ of the symmetric group.

\subsection{Fermionic (RCM) and $Q$-Schur (HM) averages}

One can substitute the bosonic matrices $Z$ in (\ref{siCompAB1}), (\ref{siCompAB2}) by the complex rectangular matrices
with Grassmann entries $\Psi$.
Then \cite{MMMZh} (see also the earlier result \cite{WWWZ})
\be
\int \S_R[A\Psi B\Psi^\dagger] \,e^{-\tr \Psi\Psi^\dagger } d^2\Psi =
(-)^{|R|}\frac{\S_R[-A ]\cdot \S_R[B ]}{ d_R}
\label{fermAB}
\ee

Surprisingly or not, a counterpart of this formula in the Hermitian case
is not yet available.
Instead, there is a counterpart of (\ref{siHermA1}) for the $Q$-Schur functions instead of the
ordinary Schur functions, which are not available in the RCM case.
There are reasons to expect that the $Q$-Schur functions are related to
fermionic averages \cite{MMZh2}, still at present the situation is that they arise
in the RCM case only, while the $Q$-Schur formulas are available only
in the Hermitian case.
According to \cite{MMkon}
\be
\left<Q_R[X]\right>:=\int Q_R[X] \,e^{-\tr X^2\Lambda} dX
= \left\{
\begin{array}{cl}
\displaystyle{{Q_{R/2}\{\Tr\Lambda^{-k}\}Q_{R/2}\{\delta_{k,1}\}\over Q_{R}\{\delta_{k,1}\}}}&\ \ \ \ \ \hbox{if }R_i
\hbox{ are all even}\cr
&\cr
0&\ \ \ \ \ \hbox{otherwise}
\end{array}
\right.
\label{QRA}
\ee
where $R/2$ denotes the Young diagram with all line lengths being half of those of $R$. This formula
implies a very nice expansion for the partition function of the cubic (original) Kontsevich model:
\be
{\int dX \exp\left(-\Tr {X^3\over 3}-\Tr X^2\Lambda \right)\over \int dX \exp\left(-\Tr X^2\Lambda\right)} = \sum_{R\in SP}   Q_R[\Lambda^{-1}]
\frac{Q_R\{\delta_{k,1}\}\cdot Q_{2R}\{\delta_{k,3}\}}{4^{|R|}\cdot Q_{2R}\{\delta_{k,1}\}}
\ee
where the sum goes over the strict partitions, i.e. partitions with all parts distinct.

\section{Quantum deformations of the measure  \label{qt}}

\subsection{$\beta$-deformation}

The previous consideration was formulated in terms of matrix integrals. Further generalizations often require the eigenvalue representation. That is, since the integral (\ref{siHerm}) involves only invariant combinations, traces of matrix powers, one can integrate over the angular variables  (or, similarly, in (\ref{siComp})) and obtain (with a properly normalized measure) \cite{Mehta}
\be\label{evH}
\int \S_R[X] \,e^{-\frac{1}{2}\tr X^2} dX \sim\int_{\mathbb{R}^N} \S_R(x)\Delta(x)^2 \,e^{-\frac{1}{2}\sum_ix_i^2} \prod_i dx_i
\ee
where $\Delta (x):=\prod_{i<j}(x_i-x_j)$ is the Vandermonde determinant, $x_i$ are eigenvalues of the matrix $X$, and the Schur function in the integrand is a symmetric function of $x_i$, or a graded polynomial of $p_k:=\sum_ix_i$: $\S_R[X]=S_R(x)=S_R\{p_k\}$.

Similarly,one can integrate over the angular variables in the RCM, which gives rise to the linear exponential instead of the Gaussian measure, and to the integration contour $\mathbb{R}_{\ge 0}$ \cite{Mehta,AMP,AMM}. For the sake of definiteness, we assume that $N_1\ge N_2$. Then,
\be\label{evC}
\int \S_R[ZZ^\dag] \,e^{-\tr ZZ^\dag} d^2Z \sim\int_{\mathbb{R}_{\ge 0}^{N_2}} \S_R(x)x^{N_1-N_2}\Delta(x)^2 \,e^{-\sum_ix_i} \prod_i dx_i
\ee
Now the natural deformation of (\ref{evH}) and (\ref{evC}) is to let the Vandermonde determinant enter with an arbitrary degree $2\beta$ (note that $\beta=1/2$ describes the real matrix integral, and $\beta=2$, the quaternion one). Then, the natural system of symmetric functions is, instead of the Schur functions, the Jack polynomials \cite{Mac} $J_R$, and the superintegrability relation looks like \cite{MPS}:
\be\label{beta}
\Big<\J_R \Big>:=
\int_{\mathbb{R}^N} \J_R(x)\Delta(x)^{2\beta} \,e^{-\frac{\beta}{2}\sum_ix_i^2} \prod_i dx_i=
 \frac{  \J_R\{\delta_{k,2}\}\cdot \J_R\{N\}}{ \J_R\{\delta_{k,1}\}}
\ee
Moreover, one can considered a little bit more general potential \cite{Zabz},
\be
\Big<\J_R \Big>:=
\int_{\mathbb{R}^N} \J_R(x)\Delta(x)^{2\beta} \,e^{-\mu\beta\sum_ix_i-\frac{\beta}{2}\sum_ix_i^2} \prod_i dx_i=
\frac{  \J_R\{\mu\delta_{k,1}+\delta_{k,2}\}\cdot \J_R\{N\}}{ \J_R\{\delta_{k,1}\}}
\label{Max2}
\ee
One can definitely consider the case $\beta=2$ in order to return to the Hermitian matrix model (\ref{siHerm}) with this more general potential.

Similarly, the superintegrability property in the complex model (\ref{evC}) is controlled by the formula \cite{Zabz}
\be
\Big<\J_R \Big>:=
\int_{\mathbb{R}_{\ge 0}^{N}} \J_R(x)\Delta(x)^{2\beta} \,e^{-\mu\beta\sum_ix_i} \prod_i x_i^ndx_i=
{\J_R\{\beta^{-1}(n+1)+N-1\} \cdot \J_R\{N\}\over \J_R\{\mu\cdot\delta_{k,1}\}}
\label{Max1}
\ee
The r.h.s. of this formula, indeed, reduces at $\beta=1$, $n=N_1-N_2$ and $\mu=1$ to the r.h.s. of (\ref{siComp}).

\subsection{$q,t$-deformations}

Further, two-parametric generalization naturally leads to the Macdonald polynomials substituting the Schur functions. This generalization introduces two parameters $q$ and $t=q^\beta$ and replaces both the Vandermonde determinant,
\be
\Delta(x)^{2\beta}\to \prod_{i\ne j}{\Big({x_i\over x_j};q\Big)_\infty\over \Big(t{x_i\over x_j};q\Big)_\infty}
\ee
and the Gaussian exponential
\be
e^{-\frac{\beta}{2}x^2}\to (q^2x^2;q^2)_\infty
\ee
or, even more generally,
\be
e^{-\frac{\beta}{2}x^2}\to (q\xi_1x;q)_\infty(q\xi_2x;q)_\infty
\ee
which reduces to $(q^2x^2;q^2)_\infty$ at $\xi_1=-\xi_2=1$.
Here $(z;q)_\infty:=\prod_{i=0}^\infty (1-zq^i)$ is the Pochhammer symbol.

With these definitions, the superintegrability relations survive \cite{MPS,Zabz}
\be
\left<\M_R\right> :=
\prod_{i=1}^N \int_{-1}^1d_qx_ix_i^{\beta(N-1)}(q\xi_1x_i;q)_\infty(q\xi_2x_i;q)_\infty\cdot  \M_R(x)\cdot \prod_{j\neq i}
\frac{\left(\frac{x_i}{x_j},q\right)_\infty}{\left(\frac{tx_i}{x_j},q\right)_\infty}
= \frac{\M_R\left\{{\xi_1^k+\xi_2^k\over 1-t^k}\right)\}\cdot \M_R\left\{\frac{1-t^{kN}}{1-t^k}\right\}}{\M_R\left\{{(-\xi_1\xi_2)^k\over 1-t^k}\right\}}
\ee
where the integral is defined to be the Jackson integral \cite{GR}.
Note that one can also consider further deformation of the RCM $\beta$-ensemble, (\ref{Max1}) with $\xi_2=0$. Then, the superintegrability property is \cite{Zabz}
\be
\left<\M_R\right> :=
\prod_{i=1}^N \int_{-1}^1d_qx_ix_i^{\beta(N-1)}(q\xi x_i;q)_\infty \cdot \M_R(x)\cdot \prod_{j\neq i}
\frac{\left(\frac{x_i}{x_j},q\right)_\infty}{\left(\frac{tx_i}{x_j},q\right)_\infty}
={\M_R\left\{{1-q^kt^{k(N-1)}\over 1-t^k}\right\} \cdot \M_R\left\{\frac{1-t^{kN}}{1-t^k}\right\}\over \M_R\left\{{\xi^k\over 1-t^k}\right\}}
\ee

\subsection{Tridiagonal model\label{trid}}

There is another interesting manifestation of the superintegrability property (\ref{beta}), that is,
\be
\prod_{i=1}^N\int_{-\infty}^\infty \!\!\!\! e^{-a_i^2/2}  da_i
\prod_{i=1}^{N-1} \int_0^\infty \!\!\!\! e^{-b_i} b_i^{\beta i-1}db_i \ \   J_R\{p_k=\tr \Phi^k\}
= \beta^{|R|\over 2}\cdot\frac{J_R\{N\}J_R\{\delta_{k,2}\}}{J_R\{\delta_{k,1}\}}
\ee
where the matrix $\Phi$ is tridiagonal:
\be
\Phi = \left(\begin{array}{ccccc}
      a_1 & b_1  & 0 & 0 & \dots \\
        1 & a_2 & b_2  & 0 & \\
      0 &  1 & a_3 & b_3  & \\
      0 & 0 &  1 & a_4 & \\
      \dots
    \end{array} \right)
\ee
The origin of this formula is the connection of this tridiagonal model with the $\beta$-ensemble of the HM \cite{DE,K,MMPtri}.

\section{Non-Gaussian  potential}

One of the ways to deform matrix model is to vary the potential. In this section, we consider two examples of non-Gaussian potentials: the monomial of higher degree, and the square of logarithm, while the potential that is a sum of two logarithms is left for sec.\ref{log}.

\subsection{Monomial non-abelian potential \label{monom}}

One can wonder what is the meaning, or, better, the generalization
of the strange factor $\S_R\{\delta_{k,2}\}$ at the r.h.s. of (\ref{siHerm}).
In order to see this, one can consider the potential which is a monomial of higher degree $s$.
According to \cite{PSh}, the superintegrability relation in this case is
\be\label{mono}
\Big< S_R \Big>_a :=
\int_{C_{s,a}^{\otimes N}} \S_R[X] \cdot e^{-\frac{1}{s}\tr X^s} dX =
\S_R\{\delta_{k,s}\}
\cdot\prod_{(\alpha,\beta)\in R}
[[N+\alpha-\beta]]_{s,0}\cdot [[N+\alpha-\beta]]_{s,a}\nn\\
\hbox{for }N = 0 \hbox{ or } a \hbox{ mod } s
\ee
Here we use the notation
\be
[[n]]_{s,a} = n\ {\rm if}\  n=a \,{\rm mod}(s) \ \  {\rm and}  \ 1 \ {\rm otherwise}
\label{sa}
\ee
Here $C_{s,a}$ is a special star-like (closed) integration contour
\be
\int_{C_{s,a}}  F(x)\ e^{-x^s/s} \,dx \ :=\ \sum_{b=1}^s e^{-2\pi i(a-1)b/s}\cdot \int_0^\infty
F\big(e^{2\pi ib/s}x\big)\ e^{-x^s/s}\, dx
\ee
which picks up only the  powers of $x$, which are equal to $a-1\, {\rm mod}\, s$, in particular,
\be
\int_{C_{s,a}} x^ke^{-x^s/s} dx=\delta_{k+1-a}^{(s)}\cdot\Gamma\left({k+1\over s}\right)
\ee
$\delta_k^{(s)}$ is defined to be 1 if $k=0$ mod $s$ and to vanish otherwise.
This makes the answer depending on an additional parameter $a=0,\ldots s-1$. The r.h.s. of (\ref{mono}) contains some factors $N+j$ from $\S_R\{N\}$, (\ref{siHerm}), that is,
those with $N+i = 0,a\, {\rm mod}(s)$, and, hence, its vanishing depends also on the value of $N$.
Note that if the condition
$N=0\ \hbox{or}\ a\ \hbox{mod}\ s$ is not satisfied, one can not define the correlator by the condition $<1>=1$ because of zeroes in the denominator.

\subsection{Square of logarithm potenital
\label{trig}}

Our next example is the potential which is a square of logarithm.
The simplest way to deal with this model is to consider exponential change of variables: $X\to e^X$. Then, one arrives at the Gaussian model again, however, with a different measure and the symmetric functions of $e^X$ instead of $X$. Let us make this substitution at the level of the eigenvalue model. Then, what happens is the trigonometric Vandermonde factor, and the superintegrability relations is
\be
\Big<\S_R \Big>:=\!
\int \S_R(e^{x_i}) \cdot \prod_{i<j}^N \sinh^2\left(\frac{x_i-x_j}{2}\right)
\prod_{i=1}^N \exp\left(-\frac{x_i^2}{2g^2}\right) dx_i =
  A^{|R|} q^{  2\varkappa_R}\cdot \S_R\left\{\frac{A^k-A^{-k}}{q^k-q^{-k}}\right\}
\ee
In fact, one can further deform the Vandermonde determinant, \cite{BEM}, and to arrive at
\be
\Big<\S_R \Big>:=
\int \S_R(e^{x_i/a}) \cdot \prod_{i<j}^N \sinh\left(\frac{x_i-x_j}{2a}\right)\sinh\left(\frac{x_i-x_j}{2b}\right)
\prod_{i=1}^N \exp\left(-\frac{x_i^2}{2g^2}\right) dx_i =
\nn \\
=  \left(A^{|R|} q^{ 2\varkappa_R}\right)^{b/a} \S_R\left\{\frac{A^k-A^{-k}}{q^k-q^{-k}}\right\}
\ee
In these formulas,  $\varkappa_R = \sum_{(\mu,\nu)\in R} (\mu-\nu)
= \frac{1}{2}\sum_\mu \lambda_\nu(\lambda_\nu-2\nu+1)$, $q:=\!\exp{\left(\frac{g^2}{2ab}\right)}$ and $A:=q^N$.

These formulas are not yet a trigonometric deformation of superintegrability, since the latter would involve symmetric functions of $x_i$, not of $e^{x_i}$.

\section{Non-Gaussian  potential: Logarithmic (Selberg) case
\label{log}}

The case of logarithmic potential is distinguished in matrix models,
because it can be interpreted either as a contribution of an additional
zero-time \cite{versus,UFN3} or as a quantum deformation.
In this case, the ordinary integrability is only slightly modified,
and the same can be (justly) expected about superintegrability.
Therefore we consider this ``simple" case separately.

\subsection{Student's distribution}

The first example is an example of the simplest model with the logarithm potential. According to \cite{MMPstud}, in this case, the superintegrability formula looks like
\be
\Big<\S_R \Big>:=\int \S_R[X]\cdot \frac{dX}{{\rm det}(1+X^2)^\alpha   } =
{1\over P_R(\alpha,N)}\cdot\frac{\S_R\{N\}\cdot \S_R\{\delta_{k,2}\}}{ \S_R\{\delta_{k,1}\}}
\label{siHermStud}
\ee
with
\be
P_R(\alpha,N) = \prod_{m=1}^{l_R} \prod_{i=1}^{\left[\frac{R_m+\delta_{m|2}}{2}\right]}
\Big(1+2\left(\alpha-N+i+[m/2]\right)\Big)
\ee
The integral (\ref{siHermStud}) is the HM. However, one can equally well consider the RCM average
\be
\int \S_R[ZZ^\dag]\cdot \frac{d^2Z}{{\rm det}(1+ZZ^\dag)^\alpha   }
\ee
which gives the same answer. In fact, until the refinement of the averages, the two models always coincide upon a proper change of the potential (and integration contours). Hence, we do not differ between them further.

\subsection{Hermitian two-logarithm model}

Instead of the Gaussian measure (\ref{siHerm}), one can consider the exponential of two logarithmic terms, which reduces, after integration over the angular variables to Selberg-type integrals \cite{Selb}, and the superintegrability in this case is given by the Kadell formulas \cite{Kad}:
\be\label{KadS}
\Big<\S_R \Big>:=
 \prod_i \int_0^1dx_ix_i^u(1-x_i)^v \S_R(x)\Delta(x)^{2}=
{\S_R\{N\}\cdot\S_R\{u+N\}\over\S_R\{u+v+2N\}}
\ee
Moreover, a similar form can be found for the Schur functions depending on inverse powers of $x_i$: making a change of variables in the multiple integral, one obtains
\be\label{rel}
\Big<\S_R\{p_{-k}\} \Big>_{u,v,N}=\Big<\S_R\{p_{k}\} \Big>_{-u-v-2N,v,N}
\ee
where $p_{-k}:=\sum_i x_i^{-k}$.

\subsection{Two-logarithm $\beta$-ensemble}

These formulas are again immediately deformed to the $\beta$-ensemble superintegrability with Jack polynomials being a proper system of symmetric functions. The average of the Jack polynomial \cite{Kad,MMSh} is
\be\label{KadJ}
\Big<\J_R \Big>:=
\prod_i \int_0^1 dx_ix_i^u(1-x_i)^v \J_R(x)\Delta(x)^{2\beta} =
{\J_R\{N\}\cdot\J_R\{\beta^{-1}u+N+\beta^{-1}-1\}\over\J_R\{\beta^{-1}(u+v+2)+2N-2\}}
\ee
while relation (\ref{rel}) becomes
\be\label{rel}
\Big<\J_R\{p_{-k}\} \Big>_{u,v,N}=\Big<\J_R\{p_{k}\} \Big>_{-u-v+2(\beta-1)-2\beta N,v,N}
\ee

\subsection{BGW model}

By a proper limit from these formulas, one can get a specific unitary matrix model, which is related to the Brezin-Gross-Witten (BGW) model \cite{BGW,AMM,PGL}. Again, in order to present the $\beta$-deformed case, one needs first to integrate over the angular variables and then to deform the degree of the Vandermonde determinant \cite{PGL}. The superintegrability relation in this case looks like \cite{MMSh}
\be\label{BGW}
\Big<\J_R \Big>:=\prod_i^N\oint_{|z_i|=1}{dz_i\over z_i}\Big|\Delta(z)\Big|^{2\beta}\mu_{N+1-\beta^{-1}}\Big\{\sum_iz_i^k\Big\}
\J_R(z)={\J_R\{\delta_{k,1}\}^2\over\J_R\{N-1+\beta^{-1}\}}
\ee
where the function $\mu_n\{p_k\}$ is defined by another $\beta$-deformed unitary integral,
\be
\mu_n\{p_k\}\Big|_{p_k=\sum_i\psi_i^k}:=\prod_i^n\oint_{|w_i|=1}{dw_i\over w_i}\Big|\Delta(w)\Big|^{2\beta}
e^{\beta \sum_i(w^+_i+\psi_iw_i)}
\ee
All other superintegrability formulas, which we discuss below are also extended to this case. However, since this model is obtained by a proper limit from model (\ref{KadJ}) \cite{PGL}, we do not write them down.

\subsection{$q,t$-deformed two-logarithm model}

As in the Gaussian case, the $\beta$-deformation of the superintegrability is immediately lifted to the $q,t$-deformation. The proper set of symmetric functions is the Macdonald polynomials, and the Selberg average of the Macdonald polynomial is \cite{Mac,MMShS,Zen}
\be\label{Kadqt}
\left<\M_R\right> :=
\prod_{i=1}^N\int_{0}^1 d_qx_i x_i^{u+\beta (N-1)}\frac{\left({x_i},q\right)_\infty}{\left({q^vx_i},q\right)_\infty}
 \cdot\M_R(x)\cdot \prod_{j\neq i}
\frac{\left(\frac{x_i}{x_j},q\right)_\infty}{\left(\frac{tx_i}{x_j},q\right)_\infty}
=
q^{|R|}t^{\nu_R}{\M_R^*\{t^N\}\cdot\M_R^*\{t^{N-1}q^{u+1}\}\over\M_R^*\{t^{2(N-1)}q^{u+v+2}\}}
\ee
where $|R|=\sum_iR_i$ is the size of the Young diagram $R$, $\nu_R:=\sum_i (i-1)R_i$, and we denoted
\be
\M_R^*\{x\}:=\M_R\left\{{1-x^k\over1-t^k}\right\}
\ee

\section{Double correlators}

\subsection{Chiral correlator in RCM}

It turns out that averages of the product of two symmetric functions of specifically correlated arguments are usually also equal to ratios of symmetric functions at special locus, i.e. superintegrability persists in these cases two. The only difference is that the average of one symmetric function is typically a ratio of two symmetric functions in the numerator and one, in the denominator, while the average of a product of two symmetric functions is typically ration of two symmetric functions both in the numerator and in the denominator, with a factor. In this section we consider examples of these averages of products of two symmetric functions.

The very first example is given by a natural complement \cite{Mac,Orlov} of eq.(\ref{siCompAB1}),
\be
\Big< \S_R[{\cal A}Z]\cdot \S_{R'}[{\cal B}Z^\dagger]\Big> =
\delta_{R,R'}\cdot \frac{\S_R[{\cal AB}]}{\S_R\{\delta_{k,1}\}}
\ee
which in no way reduces to (\ref{siCompAB1}).
In certain sense, it looks like a dual of (\ref{siCompAB1}),
with left and right hand sides exchanged.
Analogy is not full, because
the matrices ${\cal A}$ and ${\cal B}$ are now rectangular,
while $A$ and $B$ are square in (\ref{siCompAB1}).

\subsection{Hermitian two-logarithm model}

In the case of model (\ref{KadS}), there are also superintegrability formulas for the products of two Schur functions \cite{Kad1}:
\be
\Big<\S_R\{p_k+v\}\S_P\{p_k\} \Big>=C_{RP}\cdot{\S_R\{v+N\}\S_P\{u+N\}
\over\S_R\{N\}\S_P\{u+v+N\}}
\ee
where
\be
C_{RP}={\prod_{i<j}^N(R_i-i+R_j-j)(P_i-i+P_j-j)\over\prod_{i,j}^N(u+v+2N+1+R_i-i+P_j-j)}
\ee
Another type of double correlators with the superintegrability property is \cite{AFLT}
\be
\Big<\S_R\{p_{k}+v\}\S_P\{p_{-k}\} \Big>=G_{RP}(u+v+N)\cdot\S_R\{\delta_{k,1}\}\S_P\{\delta_{k,1}\}
\cdot{\S_R\{v+N\}\S_P\{N\}\over\S_R\{u+v+2N\}\S_P\{-u\}}
\ee
where we introduced the standard building block of Nekrasov functions
\be
G_{RP}(x):=\prod_{i,j\in R}(x+R_i-j+P_j^\vee-i+1)\prod_{i,j\in P}(x-P_i+j-1-R_j^\vee+i)
\ee
and $R^\vee$ denotes the conjugate Young diagram.

\subsection{$\beta$-deformed two-logarithm model}

The averages of product of the two Jack polynomials also can be found in the model (\ref{KadJ}), and are \cite{Kad1}
\be
\Big<\J_R\{p_k+\beta^{-1}(v+1)-1\}\J_P\{p_k\} \Big>=
{C_{RP}^J\over C_{\emptyset\emptyset}^J}\cdot{\J_R\{\beta^{-1}(v+1)+N-1\}\J_P\{\beta^{-1}(u+1)+N-1\}
\over\J_R\{N\}\J_P\{\beta^{-1}(u+v+2)+N-2\}}
\ee
with
\be
C_{RP}^J={\prod_{i<j}^N(R_i-\beta i+R_j-\beta j)_\beta(P_i-\beta i+P_j-\beta j)_\beta
\over\prod_{i,j}^N(u+v+2\beta N+2+R_i-\beta i+P_j-\beta j-\beta)_\beta},\ \ \ \ \ \ \ \
(x)_\beta:={\Gamma (x+\beta)\over\Gamma(x)}
\ee
and \cite{AFLT}
\be
\Big<\J_R\{p_{k}+\beta^{-1}(v+1)-1\}\J_P\{p_{-k}\} \Big>=G_{RP}^J(u+v+1+\beta (N-1))\cdot\J_R\{\beta^{-1}\delta_{k,1}\}
\J_P\{\beta^{-1}\delta_{k,1}\}\times\nn\\
\times{\J_R\{\beta^{-1}(v+1)+N-1\}\J_P\{\beta^{-1}N\}\over\J_R\{\beta^{-1}(u+v+2)+2(N-1)\}
\J_P\{-\beta^{-1}u\}}
\ee
with
\be
G_{RP}^J(x):=\prod_{i,j\in R}(x+R_i-j+\beta(P_j^\vee-i+1))\prod_{i,j\in P}(x-P_i+j-1-\beta(R_j^\vee-i))
\ee

\subsection{$q,t$-deformed two-logarithm model}

At last, there are similar formulas for the double correlators in the model (\ref{Kadqt}) \cite{MMShS}:
\be
\Big<\M_R\left\{p_k-{q^{vk}-t^kq^{-k}\over 1-t^k}\right\}\cdot\M_P\{p_k\} \Big>=q^{|R|+|P|}t^{\nu_R+\nu_P}
{C_{RP}^M\over C_{\emptyset\emptyset}^M}\cdot{\M_R^*\{t^{N-1}q^{v+1}\}\M_P^*\{t^{N-1}q^{u+1}\}
\over\M_R^*\{t^N\}\M_P^*\{t^{N-2}q^{u+v+2}\}}
\ee
with
\be
C_{RP}^J={\prod_{i<j}^N(R_i-\beta i+R_j-\beta j)_{q,t}(P_i-\beta i+P_j-\beta j)_{q,t}
\over\prod_{i,j}^N(u+v+2\beta N+2+R_i-\beta i+P_j-\beta j-\beta)_{q,t}},\ \ \ \ \ \ \ \
(x)_{q,t}:={(q^x;q)_\infty\over (q^xt;q)_\infty}
\ee

\section{Conclusion}

To conclude, we reviewed the progress achieved in extending the
superintegrability property in various directions.
Despite unexpectedly slow, this progress is quite impressive:
it is clear that the phenomenon is not accidental, and is reasonably general.
The main advantage of our approach is a conceptual reference
to some additional symmetry manifested by superintegrability,
which, however, remains to be better explained and interpreted.
Also important is relation to multi-diagonal formulas\footnote{
To avoid possible confusion, here this term, say, {\it tri-diagonal},
is used in an absolutely different sense than in sec.\ref{trid}.
} for
Hurwitz $\tau$-functions \cite{AMMN},
which are getting more and more applications in different branches
of mathematical physics.
This direction originates from the Natanzon-Orlov
generalization of (\ref{siComp}) \cite{NO,MMMZh},
it is rather new and deserves explaining more details
than other subjects.
We  consider it in a separate paper \cite{MMMZh2}.
We hope that this summary will attract more researchers into
the field, which is relatively simple and does not require
much of special knowledge.
This would lead to new and faster progress,
both conceptual and technical.

\section*{Acknowledgements}

This work was supported by the Russian Science Foundation (Grant No.21-12-00400).


\begin{thebibliography}{12}

\bibitem{MM} A.~Mironov, A.~Morozov,
  Phys.\ Lett.\ {\bf B771} (2017) 503,
arXiv:1705.00976

\bibitem{DiF} P.~Di Francesco, C.~Itzykson, J.~B.~Zuber,
  Commun.\ Math.\ Phys.\  {\bf 151} (1993) 193,
  hep-th/9206090

\bibitem{IdF} P.~Di Francesco, C.~Itzykson,
Ann. Inst. H. Poincare Phys. Theor. \textbf{59} (1993) 117-140,
hep-th/9212108

\bibitem{Ivan1} I.~K.~Kostov, M.~Staudacher,
Phys. Lett. \textbf{B394} (1997) 75-81,
hep-th/9611011

\bibitem{Ivan2} I.~K.~Kostov, M.~Staudacher, T.~Wynter,
Commun. Math. Phys. \textbf{191} (1998) 283-298,
hep-th/9703189

\bibitem{Orlov} A. Orlov, 
Int. J. Mod. Phys. {\bf A19, supp02} (2004) 276-293, nlin/0209063

\bibitem{MMSh} A.~Mironov, A.~Morozov, S.~Shakirov,
JHEP \textbf{02} (2011) 067,
arXiv:1012.3137

\bibitem{MMShS} A.~Mironov, A.~Morozov, S.~Shakirov, A.~Smirnov,
Nucl. Phys. \textbf{B855} (2012) 128-151,
arXiv:1105.0948

\bibitem{AMMN} A. Alexandrov, A. Mironov, A. Morozov, S. Natanzon,
JHEP 11 (2014) 080,  arXiv:1405.1395

\bibitem{Orlov2}  S. Natanzon, A. Orlov, arXiv:1407.8323

\bibitem{PSh} C.~Cordova, B.~Heidenreich, A.~Popolitov, S.~Shakirov,
  Commun.\ Math.\ Phys.\  {\bf 361} (2018)   1235,
  arXiv:1611.03142

  \bibitem{IMM} H.~Itoyama, A.~Mironov, A.~Morozov,
  JHEP {\bf 1706} (2017) 115,
arXiv:1704.08648

  \bibitem{MMten} A.~Mironov, A.~Morozov,
  Phys.\ Lett.\ {\bf B774} (2017) 210,
arXiv:1706.03667

\bibitem{MPS} A.~Morozov, A.~Popolitov and S.~Shakirov,
  Phys.\ Lett. {\bf B784} (2018) 342,
  arXiv:1803.11401

    \bibitem{MMsum} A.~Mironov, A.~Morozov,
  JHEP {\bf 1808} (2018) 163,
arXiv:1807.02409

\bibitem{MMell}   A.~Mironov, A.~Morozov,
Phys. Lett. \textbf{B816} (2021), 136196,
arXiv:2011.01762; ibid.,
136221,
arXiv:2011.02855

\bibitem{MMkon} A.~Mironov, A.~Morozov,
Eur. Phys. J. \textbf{C81} (2021) 270,
arXiv:2011.12917

\bibitem{MMl} A.Mironov, A.Morozov,  Phys.Lett. {\bf B816} (2021) 136268,  arXiv:2102.01473

\bibitem{Zabz} L.~Cassia, R.~Lodin, M.~Zabzine,
JHEP \textbf{10} (2020) 126,
arXiv:2007.10354

\bibitem{MMPstud} A.~Mironov, A.~Morozov, A.~Popolitov,
Phys. Lett. \textbf{B824} (2022) 136833,
arXiv:2107.13381

\bibitem{WWWZ} L.~Y.~Wang, R.~Wang, K.~Wu, W.~Z.~Zhao,
Nucl. Phys. \textbf{B973} (2021) 115612,
arXiv:2110.14269

\bibitem{MMMZh} A.~Mironov, V.~Mishnyakov, A.~Morozov, A.~Zhabin,
arXiv:2112.11371

\bibitem{Kaz} V.A. Kazakov, M. Staudacher, T. Wynter, 
hep-th/9601153, 1995 Carg\`ese Proceedings

\bibitem{Ramg} S. Corley, A. Jevicki, S. Ramgoolam, Adv.Theor.Math.Phys. {\bf 5} (2002) 809-839, hep-th/0111222

\bibitem{KPSS} C. Kristjansen, J. Plefka, G. W. Semenoff, M. Staudacher, Nucl.Phys. {\bf B643} (2002) 3-30, hep-th/0205033

\bibitem{BEM} M. Tierz, Mod. Phys. Lett. A19 (2004) 1365-1378, hep-th/0212128\\
A. Brini, B. Eynard, M. Mari\~no, Annales Henri Poincar\'e. Vol. 13. No. 8. SP Birkh\"{a}user Verlag Basel, 2012, arXiv:1105.2012

\bibitem{MKR} R. de Mello Koch, S. Ramgoolam, arXiv:1002.1634

\bibitem{Mac} I.G. Macdonald,
{\it Symmetric functions and Hall polynomials}, Second Edition, Oxford University Press,
1995

\bibitem{Fulton} W. Fulton, {\sl Young tableaux: with applications to representation theory and geometry},
LMS, 1997

\bibitem{MMN} A.~Mironov, A.~Morozov, S.~Natanzon,
Theor. Math. Phys. \textbf{166} (2011) 1-22,
arXiv:0904.4227;
J. Geom. Phys. \textbf{62} (2012) 148-155,
arXiv:1012.0433

\bibitem{UFN3} A. Morozov,
Phys.Usp.(UFN) {\bf 37} (1994) 1;
hep-th/9502091; hep-th/0502010\\
A. Mironov, Int.J.Mod.Phys. {\bf A9} (1994) 4355; Phys.Part.Nucl.
{\bf 33} (2002) 537; hep-th/9409190

\bibitem{NO}
S.Natanzon, A.Orlov,
Teor. Mat. Fiz. \textbf{204} (2020) 396-429
[erratum: Theor. Math. Phys. \textbf{205} (2020) 1546],
arXiv:2006.07396

\bibitem{MMZh2} A.~Mironov, A.~Morozov, A.~Zhabin,
arXiv:2111.05776

\bibitem{Mehta} J. Ginibre,
J. Math. Phys. {\bf 6} (1965) 440\\
M.L. Mehta, {\it Random Matrices,} 2.ed.,
Academic Press, 1990

\bibitem{AMP} A.~Anderson, R.~C.~Myers, V.~Periwal,
Phys. Lett. \textbf{B254} (1991) 89-93

\bibitem{AMM} A.~Alexandrov, A.~Mironov and A.~Morozov,
JHEP \textbf{12} (2009) 053,
arXiv:0906.3305

\bibitem{GR} G. Gasper, M. Rahman, {\it Basic hypergeometric series},
Cambridge University Press, 1990

\bibitem{DE} I. Dumitriu, A. Edelman,
J. Math. Phys. 43(11) (2002) 5830-5847,
math-ph/0206043

\bibitem{K} R. Kozhan,
in: {\it Operator Theory: Advances and Applications}, {\bf vol.276} (2020) 434–447,
arXiv:1801.05749

\bibitem{MMPtri} A.~Mironov, A.~Morozov, A.~Popolitov,
arXiv:2110.14005

\bibitem{versus} S.~Kharchev, A.~Marshakov, A.~Mironov, A.~Morozov,
Nucl. Phys. \textbf{B397} (1993) 339-378,
hep-th/9203043

\bibitem{Selb} A. Selberg, 
Norsk. Mat. Tidsskr. {\bf 24} (1944) 71-78

\bibitem{Kad} J. Kaneko,
SIAM.J.Math.Anal. {\bf 24} (1993) 1086-1110 \\
K.W.J. Kadell,
Adv.Math. {\bf 130} (1997) 33-102

\bibitem{BGW} A.~Mironov, A.~Morozov, G.~W.~Semenoff,
Int. J. Mod. Phys. \textbf{A11} (1996) 5031-5080,
arXiv:hep-th/9404005

\bibitem{PGL} A.~Mironov, A.~Morozov and S.~Shakirov,
JHEP \textbf{03} (2011) 102,
arXiv:1011.3481

\bibitem{Zen} Y.~Zenkevich,
JHEP \textbf{05} (2015) 131,
arXiv:1412.8592

\bibitem{Kad1} K.W.J. Kadell,
Compositio Math. {\bf 87} (1993) 5-43

\bibitem{AFLT} V.~A.~Alba, V.~A.~Fateev, A.~V.~Litvinov, G.~M.~Tarnopolskiy,
Lett. Math. Phys. \textbf{98} (2011) 33-64,
arXiv:1012.1312

\bibitem{MMMZh2} A. Mironov, V. Mishnyakov, A. Morozov, A. Zhabin, to appear

\end{thebibliography}
\end{document}